\documentclass{aastex}
\usepackage{emulateapj5}

\shortauthors{Clarke, Kronberg, \& B\"ohringer}
\shorttitle{Intracluster Magnetic Fields}

\newcommand{\radm}{${\rm rad\ m^{-2}}$}
\newcommand{\revm}{Rev.\ Mod.\ Phys.}
\newcommand{\revma}{Rev.\ Mod.\ Astron.}

\begin{document}

\title{A New Radio $-$ X-Ray Probe of Galaxy Cluster Magnetic Fields}

\author{T.~E. Clarke\altaffilmark{1}}
\affil{Department of Astronomy, University of Toronto, 60 St. George St., Toronto, ON M5S 3H8  Canada}
\email{tclarke@aoc.nrao.edu}

\author{P.~P. Kronberg}
\affil{Department of Physics, University of Toronto, 60 St. George St., Toronto, ON M5S 1A7  Canada}

\and

\author{Hans B\"ohringer}
\affil{Max-Planck-Institut f\"ur extraterrestrische Physik, D-85740
Garching, Germany}

\altaffiltext{1}{present address: NRAO, 1003 Lopezville Rd., Socorro,
N.M.  87801 USA}

\begin{abstract}

Results are presented of a new VLA$-$ROSAT study that probes the
magnetic field strength and distribution over a sample of 16
``normal'' low redshift ($z \leq 0.1$) galaxy clusters. The clusters
span two orders of magnitude in X-ray luminosity, and were selected to 
be free of (unusual) strong radio cluster halos, and
widespread cooling flows. Consistent with these criteria, most
clusters show a relaxed X-ray morphology and little or no evidence for
recent merger activity. 

Analysis of the rotation measure (RM) data shows cluster-generated
Faraday RM excess out to ${\sim} 0.5\, h_{75}^{-1}$ Mpc from cluster
centers. The results, combined with RM imaging of cluster-embedded
sources and ROSAT X-ray profiles indicates that the hot intergalactic
gas within these ``normal'' clusters is permeated with a high filling
factor by magnetic fields at levels of $<|B|>_{icm}$ = 5$-$10 $(\ell
/10\ {\rm kpc})^{-1/2}\,h_{75}^{1/2}\ \mu $G, where $\ell$ is the
field correlation length. These results lead to a global estimate of
the total magnetic energy in clusters, and give new insight into the
ultimate energy origin, which is likely gravitational. These results
also shed some light on the cluster evolutionary conditions that
existed at the onset of cooling flows.

\end{abstract}

Subject headings: galaxies: clusters: general --- magnetic fields ---
polarization --- radio continuum: galaxies --- X-rays: general

\section{INTRODUCTION}

Determinations of magnetic field strengths in the intracluster medium
over the past decade have revealed fields of unanticipated strength in
some clusters. Additionally, these studies have raised apparent
contradictions between field strengths (or limits) obtained from
different observational methods. In order to understand the physical
conditions in the intracluster medium, it is important to understand the
origin(s) of cluster fields, and the astrophysical processes that
relate them to other energy constituents of the cluster gas.

Estimates of magnetic field strengths in the intracluster medium (ICM)
can be determined from measurements of Faraday rotation through the
intracluster gas, and an independent measurement of the ICM thermal
electron density, $n_{e}$. Faraday rotation is given by:
\begin{equation}
{\rm RM}=\frac{\Delta \chi}{\Delta \lambda^2}=811.9\int_0^L n_{e} B_{\|}d\ell
\ \ {\rm rad\ m^{-2}},
\label{eqn:rm}
\end{equation}
where $\chi$ is the position angle of the linearly polarized radiation
at wavelength $\lambda$, $n_{e}$ is the thermal electron density in
${\rm cm^{-3}}$, $B_{\|}$ is the line of sight magnetic field strength
in $\mu$G, and $L$ is the path length in kpc. The thermal electron
density in a cluster can be determined from X-ray surface brightness
profiles of the hot (T${\rm \sim 10^8}$ K), diffuse (${n_{e}}\sim
10^{-3}\,h_{75}^{1/2}\ {\rm cm^{-3}}$) gas \citep{hans95} which fills
the cluster potential.

The first study of background RMs over a {\it single} cluster
\citep{kk90} was a targeted set of deep VLA observations of 18 sources
close in angular position to the Coma cluster (which has no cooling
flow, but a strong synchrotron halo). The result of the study was $ <|B|>_{icm}$ =
2.5 $(\ell /10\ {\rm kpc})^{-1/2}\,h_{75}^{1/2}\ \mu $G where $\ell$ is
the B-reversal scale. In a subsequent study by
\citet{feretti95} the discovery of smaller $\ell$ scales down to 1 kpc
raised this estimate to $\sim 7.2\,h_{75}^{1/2}\mu$G for Coma.

Studies of a large number of galaxy clusters with many RM probes per
cluster are currently unfeasible given the sensitivity limits of
available radio telescopes, combined with the small angular size of
more distant clusters. This situation can be circumvented by obtaining
RM probes through a sample of clusters, each having typically only one
or two (bright) polarized radio sources. The consensus of these
studies \citep{ld82, kim91, gr93} is that cluster cores have a
detectable component of RM, and many have field strengths at the
microgauss level. Note that the Faraday study by \citet{hoe89} does
not find evidence for intracluster magnetic fields. This discrepancy
is, however, likely due to the combination of small statistics and
large impact parameters in their study.

For a few clusters with extensive cooling flows, high resolution
Faraday rotation measure mapping of extended sources that are embedded
within the cooling flow zones have, in combination with X-ray data,
produced magnetic field strength estimates of 10--40 $\mu$G, ordered
on scales varying from 100 -- 0.5 kpc (see Taylor, Allen, \& Fabian 1999,
and references therein). 

Lower limits to ICM magnetic field strengths in the range 0.1 -- 1
$\mu$G level have been suggested from recent detections of both excess
(over thermal) extreme ultraviolet (EUV) and hard X-ray (HEX) emission
in some clusters. These field values would seem to {\it prima facie}
contradict the much higher values above {\it if} the EUV excess
emission is interpreted as inverse Compton (IC) scattering of $\sim$
100 MeV electrons \citep{rug94}. The EUV detections \citep{bowy1} in
particular apply to only a few clusters having widespread extended
synchrotron emission and/or they occur near to the cooling flow zone
(as in M87). Spatial differentiation of high and low magnetic field
regions in the ICM avoids the apparent contradiction by allowing the
synchrotron and EUV IC emission to originate in low field regions,
while high field regions, where the synchrotron energy loss time is
short, would provide the major contribution to the Faraday rotation
measures \citep{ensslin1}. Further, EUV emission also appears not to
be cluster wide, but concentrated to central sub-regions, and hence
does not spatially correspond to the wider regions probed by the X-ray
emission and Faraday RM measurements. The HEX excess in clusters can
be plausibly understood as bremsstrahlung from a (probably shock
heated) population of suprathermal electrons
\citep{ensslin1,dogiel1,sarazin00}. Thus, on balance, it appears that
spatial differentiation of field regions and/or emission mechanisms
other than IC scattering of CMB photons are more plausible where
apparent contradictions in magnetic field estimates arise.

This paper concentrates on what we shall term ``normal'' Abell clusters
{\it i.e.\ }those which have neither {\it widespread} cooling flows nor 
strong synchrotron halos. The observations are aimed at estimating the
strength and spatial extent of intracluster medium magnetic fields for
a relatively homogeneous sample of 16 Abell clusters. Throughout this 
{\it Letter}, we adopt $H_o=75\,h_{75}\ {\rm km\ s^{-1}\ 
Mpc^{-1}}$, and $q_o$=0.5.
 
\section{SELECTION OF THE CLUSTERS}
\label{sect:selection}

Each target cluster in our sample was required to have bright (${\rm
L_x} > 5 \times 10^{42}$\,$h_{75}^{-2}\ {\rm ergs\ s^{-1}}$), extended
X-ray emission in ROSAT observations. Therefore, the majority of the
target galaxy clusters in our sample fall within the low redshift ($z
\leq 0.1$) part of the X-ray-brightest Abell-type clusters of galaxies
\citep{xbacs} sample. The XBACs clusters are limited to high Galactic
latitudes (${\rm |b|\geq 20\degr }$), low redshifts ($z \leq 0.2$),
and ROSAT 0.1--2.4 keV band X-ray fluxes above ${\rm 5.0\times
10^{-12}\ ergs\ cm^{-2}\ s^{-1}}$. A further selection constraint was
that each cluster was required to have at least one linearly polarized
radio source viewed through the X-ray emitting gas. Such radio sources
are referred to as the {\it cluster} sample. A second set of polarized
({\it control}) radio sources viewed, in projection, outside the
boundary of the X-ray emission was also selected for each target
galaxy cluster. All polarized radio targets ({\it cluster} and {\it
control}) were selected from the NRAO VLA Sky Survey \citep{nvss}
data.

Galaxy cluster selection was further constrained such that the
sightlines of the target radio sources probed, collectively, the
largest possible range of impact parameters. The sources also had to
have sufficient polarized flux density that follow-up polarimetry
could be undertaken in short integrations ($\sim$ 5 minutes) at the
VLA\footnote{The National Radio Astronomy Observatory is a facility of
the National Science Foundation operated under cooperative agreement
by Associated Universities, Inc}. Specifically the inclusion criterion
required $I_{1.4}>$ 100 mJy and 1.4 GHz polarization, $m_{1.4}$,
greater than 1\% in the NVSS survey. A complete description of the
galaxy cluster and radio source samples will be presented in Clarke,
Kronberg, \& B\"ohringer (in preparation). 

The final sample consists of 16 Abell clusters, 13
of which are members of the XBACs sample. The three non-XBACs clusters
fall slightly below the flux limit for inclusion in the XBACs
sample. Our 16 cluster sample was reduced from an original 24 that
fell within the above criterion as severe radio frequency
interference, which is endemic to some of the crucial (for RM) 20 cm
bands, reduced the reliability of some RMs. To optimize data quality,
we reduced the final sample to 16, since even this smaller number was
statistically adequate.

It is important to mention {\it ab initio} two types of systematic
bias that might result from the above selection criteria. First, the
condition $m_{1.4}>$ 1\% may preferentially select against regions of
very high Faraday rotation, whose signature would be low
polarization at the longer radio wavelengths. This could have caused
some high RMs, but not low RMs, to have been
missed in our cluster sample. This would statistically
understate the clusters' true rotation measure distribution, and hence
magnetic field strengths. Second, the very innermost regions of the
cluster cores, which in some cases may have a cooling flow, will have been
missed because of their small angular cross sections. We do not
consider this second form of bias to be serious, since this
investigation is targeted to cluster volumes that do {\it not} have
strong cooling flows. 

\section{OBSERVATIONS AND ANALYSIS OF THE DATA}

\subsection{Radio}

Target radio sources selected from the NVSS survey were re-observed
with the VLA at four to six wavelengths within the 20 cm and 6 cm
bands. These wavelengths were selected to provide Faraday rotation
measures that are unambiguous within the range ${\rm |RM| \leq
2600}$ \radm. The observations were undertaken in October and
December 1995, August 1996, and September 1997 in the VLA's B, D, and
CS configurations respectively.

The radio data were reduced within the NRAO AIPS package following the
standard Fourier transform, deconvolve, and restore method. In
addition, self-calibration was applied to each source to further
reduce the effects of phase fluctuations. Images in the Stokes I, Q,
and U parameters were produced for each source at each of at least
four wavelengths.

\subsection{X-ray}

X-ray observations of each galaxy cluster were retrieved from the
ROSAT Data Archive\footnote{The ROSAT Data Archive is maintained by
the Max-Planck-Institut f\"ur extraterrestrische Physik (MPE) at
Garching, Germany.}. Thirteen of the target clusters were in the
Pointed Observation archive while the data for the remaining three
clusters were extracted from the ROSAT All-Sky Survey archive. All
extracted X-ray data were taken using one of ROSAT's Position
Sensitive Proportional Counters (PSPCs) which have moderate angular
resolution (FWHM $\sim$ 25\arcsec) and are sensitive to photons in
range of 0.1 -- 2.4 keV.

The ROSAT X-ray data were reduced using the Extended X-ray Scientific
Analysis Software \citep{exsas} package under the European Southern
Observatory's (ESO's) Munich Image Data Analysis System (MIDAS). A
radial X-ray surface brightness profile was determined for each
cluster by integrating the PSPC photon events over concentric annuli
of 15\arcsec\ and 30\arcsec\ for the ROSAT Pointed and RASS
Observations respectively. The surface brightness profiles were fit
with a hydrostatic isothermal model \citep{sarazin86}.

\section{RESULTS}
\label{results}

The observed Faraday rotation measure of an individual source is an
algebraic sum of Faraday contributions due to our Galaxy, the cluster,
any source-intrinsic component, and the general IGM. The latter three
are usually small, and the Galactic RM contribution was statistically
removed by subtracting the mean RM over all non-cluster sources within
10$^\circ$ of the cluster center from that of the radio probe in
question. Due to the high Galactic latitudes of the target clusters,
the mean Galactic contribution in the present sample is fairly small,
on average 9.5 rad ${\rm m^{-2}}$. The cluster radio sources and
associated Galaxy-corrected RMs are listed in Table~\ref{table}.

In Figure~\ref{RRM} we plot the residual rotation measure of the radio
sources as a function of cluster impact parameter in kiloparsecs. This
figure displays a clear Faraday excess in radio sources viewed through
the X-ray emitting ICM (open points) as compared to those viewed
beyond the detectable edge of the thermal cluster gas. The RM
distribution of the {\it control} sources has a width of 15 \radm\
while that of the {\it cluster} sources viewed through the ICM is much
broader at 114 \radm.  The Kolmogorov-Smirnov test rejects the null
hypothesis that the two samples were drawn from the same population
with a confidence level of 99.5\%. This confirms the detection of an
intracluster Faraday rotating medium.

Determination of magnetic field strengths at various impact parameters
within our cluster sample requires some assumption about the field
topology along the line of sight to the radio probe. The simplest
model of the ICM magnetic fields is a ``uniform slab'' in which the
magnetic field has constant strength and direction through the entire
cluster. Using this model and the X-ray determined electron densities,
Equation~\ref{eqn:rm} yields magnetic field strengths between $\sim$
0.5 -- 3.0$\,h_{75}^{1/2}$ $\mu$G across the Faraday sample. More
realistically, there are reversals along the line of sight to the
radio source probe. In the simple case of an intracluster medium
composed of $\mathcal{N}$ cells of uniform size and field strength,
but random field directions, the field in an individual cell increases
as $\sqrt{\mathcal{N}}$ over the uniform slab estimate. Using a simple
tangled cell model with a constant coherence length, $\ell$ =
10\,$h_{75}^{-1}$ kpc, yields an average intracluster magnetic field
strength estimate of $\sim 5(\ell /10\ {\rm
kpc})^{-1/2}\,h_{75}^{1/2}\ \mu$G. This coherence length $\ell$,
estimated from RM images of three extended radio sources in our sample
(see below), is limited by the resolution of our images. Because
$<\ell(r)>$ may systematically change ({\it e.g.}  increase with $r$)
this ``global'' average field value could increase to values as high
as 10$\,h_{75}^{1/2}\ \mu G$ near the cluster cores.

The distribution of excess RM's appears to cutoff close to the
observed X-ray outer boundary. This result, though interesting and
new, requires more detailed X-ray and radio data to understand the
variation of the magnetic to thermal energy density
${\varepsilon_{B}(r)}$/${\varepsilon_{th}(r)}$ throughout the ICM.

A striking effect, seen in Figure~\ref{RRM}, is the virtual exclusion
of small RMs at $r\, \leq \,500\,h_{75}^{-1}\rm \,kpc$. This suggests
that the RM filling factor in normal galaxy clusters is very high. An
independent, quantitative estimate of the RM filling factor is
provided from analysis of multi-frequency polarization images of three
extended radio sources, 0039+212, 0056-013, and 1650+815 (J2000) that
are embedded within three of our clusters (Abell 75, 119, and 2247
respectively). These sources project a combined area of $2.5 \times
10^{4}\, \rm kpc^{2}$, and consist of 3 sets of 30, 48, and 24
contiguous independent RM sightlines, each of which has approximately
the same areal cross-section: $\sim(5\, \rm kpc)^{2}$. The individual
RM histograms across the three sources are consistent with normal
distributions with means of -60, -144, and -97 $\rm rad\, m^{-2}$. We
find that 95\% of the sightlines within this subset of three clusters
have RMs significantly higher than the dispersion for non-cluster
sightlines, $\rm |RM|\, = 15 \, \rm rad\, m^{-2}$. This implies that
the areal filling factor of magnetic fields is {\it at least} 95\% in
the ICM. This strongly suggests that these enhanced magnetic field
levels permeate the clusters with a high filling factor, since within
cell sizes down to a resolution of 10 kpc almost no ray passing
through the ICM escapes some magnetized region.

\section{DISCUSSION}
\label{discussion}

These results confirm the widespread existence of magnetic fields in
the central regions of non cooling flow clusters. The
cluster-enhanced RM can generally be traced out to the periphery of
the ROSAT-detectable ICM X-ray emission. The rotation measure
distribution across our cluster sample drops from $\sim$ 200 ${\rm
rad\ m^{-2}}$ in the central regions (which may be an underestimate,
see \S 2) to the background level of $\sim$ 15 ${\rm rad\ m^{-2}}$ at
large radii.

Our new measurements of (1) the RM, (2) the intracluster electron
density, (3) the magnetic field volume filling factor, and (4) the
average tangling scale of the field enable us to estimate, even if
only crudely, an important physical quantity $-$ the total energy in
the ICM magnetic field for ``normal'', non-merging, relaxed
clusters. For a 5$\,h_{75}^{1/2}\ \mu $G magnetic field in the inner
500\,$h_{75}^{-1}$ kpc sphere, the total magnetic energy is $E_{B} =
1.5 \times 10^{61}({\frac{r}{500\ {\rm kpc}}})^{3}(\frac{B} {5\ \mu
{\rm G}})^{2}$\,$h_{75}^{-2}\ \rm ergs$. The magnetic energy content
of the ICM can then be compared to the total thermal energy content in
the same cluster volume. The latter is (again taking constant values
within a fiducial radius that is close to both the RM cutoff radius in
Figure~\ref{RRM} and the X-ray radius) $E_{th} = 6.4 \times
10^{62}\,n_{e}\left ({\frac{r}{500\ {\rm kpc}}}\right )^{3} \left
(\frac{T}{10^{8}\ {\rm K}}\right ) \,h_{75}^{-2}\ {\rm ergs}$, where
$n_{e}$ is the intracluster electron density in units of
$10^{-3}\,h_{75}^{1/2}\ {\rm cm^{-3}}$. The ratio $
\frac{E_{B}}{E_{th}}$ is of order 2.5\%, and the possibility that
$|B|$ may be underestimated due to limited radio resolution makes it
possible that $ \frac{E_{B}}{E_{th}}$ could be even higher. Even at
the lower bound of 2.5\% the ratio suggests that the magnetic energy
provides a non-negligible fraction of the energy budget of the ICM.

We can now compare this approximate magnetic energy estimate with
other sources of energy that are relevant for a cluster: the
thermonuclear (stellar) energy released in all cluster-member
galaxies, the gravitational binding energy released from AGN/accretion
disks, the cluster gas binding energy, and the energy associated with
past merger events. The available energy from stellar sources cannot
be much greater than $\sim 10^{62}$ ergs \citep{voelk00}, {\it i.e.}
of comparable magnitude, and is thus insufficient to maintain the
magnetic fields, barring a very high energy conversion
efficiency. This means that thermonuclear energy can be ruled out as
the primary source of intracluster magnetic field energy. It must
therefore be ultimately derived from gravity. A single powerful
AGN/accretion disk can be expected to inject approximately $10^{61}$
ergs over its lifetime into the ICM. Comparing the lifetime of the
radio source with the cluster lifetime we expect $10^2$ sources to
have injected a total of $\sim 10^{62} - 10^{63}$ ergs into the
ICM. This makes AGN/accretion disks an attractive possible source of
the ICM field energy, as has been suggested by
\citet{colgate99}. Other sources of ICM magnetic energy are the
gravitational binding energy of the cluster gas, which is of order
$10^{64}$ ergs, and the energy associated with a past cluster merger
event, which is of order $10^{63-64}$ ergs. This suggests that the
magnetic energy possibly ``taps into'' a combination of the
gravitational energy released in AGN/accretion disks over the
lifetimes of clusters, and shearing and shocks associated with larger
scale infall of matter as the clusters evolve.

Given that the cooling flows represent a late stage of cluster
evolution and our ``global'' value of 5 $\,h_{75}^{1/2} \mu G$ for ICM
zones within the clusters, but outside of cooling flow zones, we
conclude that the cooling flow develops out of a medium whose field
strength is already a significant fraction of what is seen in the
cooling flow zones.

\begin{acknowledgements}

P.P.K.\ acknowledges the support of the Natural Sciences and
Engineering Research Council of Canada (NSERC) and the support of a
Killam Fellowship, and T.E.C.\ is grateful for the support of NSERC,
OGS, and I.O.D.E.\ scholarships, and the Sumner Fellowship. We
acknowledge beneficial discussions with Stirling Colgate,  Jean Eilek, Torsten En\ss lin, Jim Felten, and Greg Taylor, and we thank the
referee for helpful comments.
   
\end{acknowledgements}

\clearpage

\begin{table}
\begin{center}
\caption{Cluster Sample Radio Sources \label{table}}
\begin{tabular}{lrcrrc} \tableline\tableline
Source & Abell & ${\rm L_x}$ \tablenotemark{a} & b \tablenotemark{b} & RM \tablenotemark{c} & $\pm$ RM\\
(J2000) & & ($ 10^{44}\ {\rm ergs\ s^{-1}}$) & (kpc) & \multicolumn{2}{c}{(rad ${\rm m^{-2}}$)} \\ \tableline
0039+212 & 75 & 0.26 & 100 & $-$62.62 & 9.03 \\
0040+212 & 75 & 0.26 & 480 & $-$34.00 & 10.95 \\
0042$-$092 & 85 & 3.96 & 960 & 4.88 & 9.77 \\
0056$-$013 & 119 & 1.55 & 280 & $-$149.16 & 10.68 \\
0057$-$013 & 119 & 1.55 & 1020 & 14.02 & 10.17 \\
0126$-$013a & 194 & 0.06 & 140 & 94.22 & 12.25 \\
0154+364 & 262 & 0.28 & 570 & $-$202.36 & 4.31 \\
0245+368 & 376 & 0.57 & 630 & $-$48.06 & 5.92 \\
0257+130 & 399 & 2.60 & 280 & $-$185.89 & 6.47 \\
0318+419 & 426 & 6.36 & 550 & 6.17 & 15.40 \\
0316+412 & 426 & 6.36 & 740 & 74.94 & 9.26 \\
0434$-$131 & 496 & 1.52 & 310 & 52.91 & 6.69 \\
0434$-$133 & 496 & 1.52 & 360 & 35.89 & 6.73 \\
0709+486 & 569 & 0.02 & 3 & $-$229.74 & 7.75 \\
0909$-$093 & 754 & 3.98 & 1010 & $-$20.47 & 9.16 \\
0908$-$100 & 754 & 3.98 & 1460 & 12.35 & 9.43 \\
0919+334 & 779 & 0.07 & 560 & 16.55 & 8.38 \\
1037$-$270 & 1060 & 0.23 & 410 & 9.38 & 6.18 \\
1037$-$281 & 1060 & 0.23 & 510 & 25.21 & 8.19 \\
1039$-$273 & 1060 & 0.23 & 530 & 103.99 & 6.33 \\
1039$-$272 & 1060 & 0.23 & 620 & 103.73 & 11.51 \\
1036$-$267 & 1060 & 0.23 & 640 & 25.76 & 16.23 \\
1133+489a & 1314 & 0.20 & 270 & 68.05 & 6.42 \\
1133+490 & 1314 & 0.20 & 310 & $-$50.38 & 6.40 \\
1145+196 & 1367 & 0.68 & 170 & 257.46 & 11.73 \\
1650+815a & 2247 & 0.06 & 200 & $-$127.36 & 8.56 \\ 
1650+815b & 2247 & 0.06 & 270 & $-$131.75 & 8.56 \\ \tableline
\end{tabular} 
\tablenotetext{a}{X-ray luminosity in the ROSAT 0.1 -- 2.4 keV band determined from the current work.}
\tablenotetext{b}{Cluster-centric impact parameter of radio source.}
\tablenotetext{c}{Galaxy-corrected rotation measure.}
\end{center}
\end{table}

\clearpage

\plotone{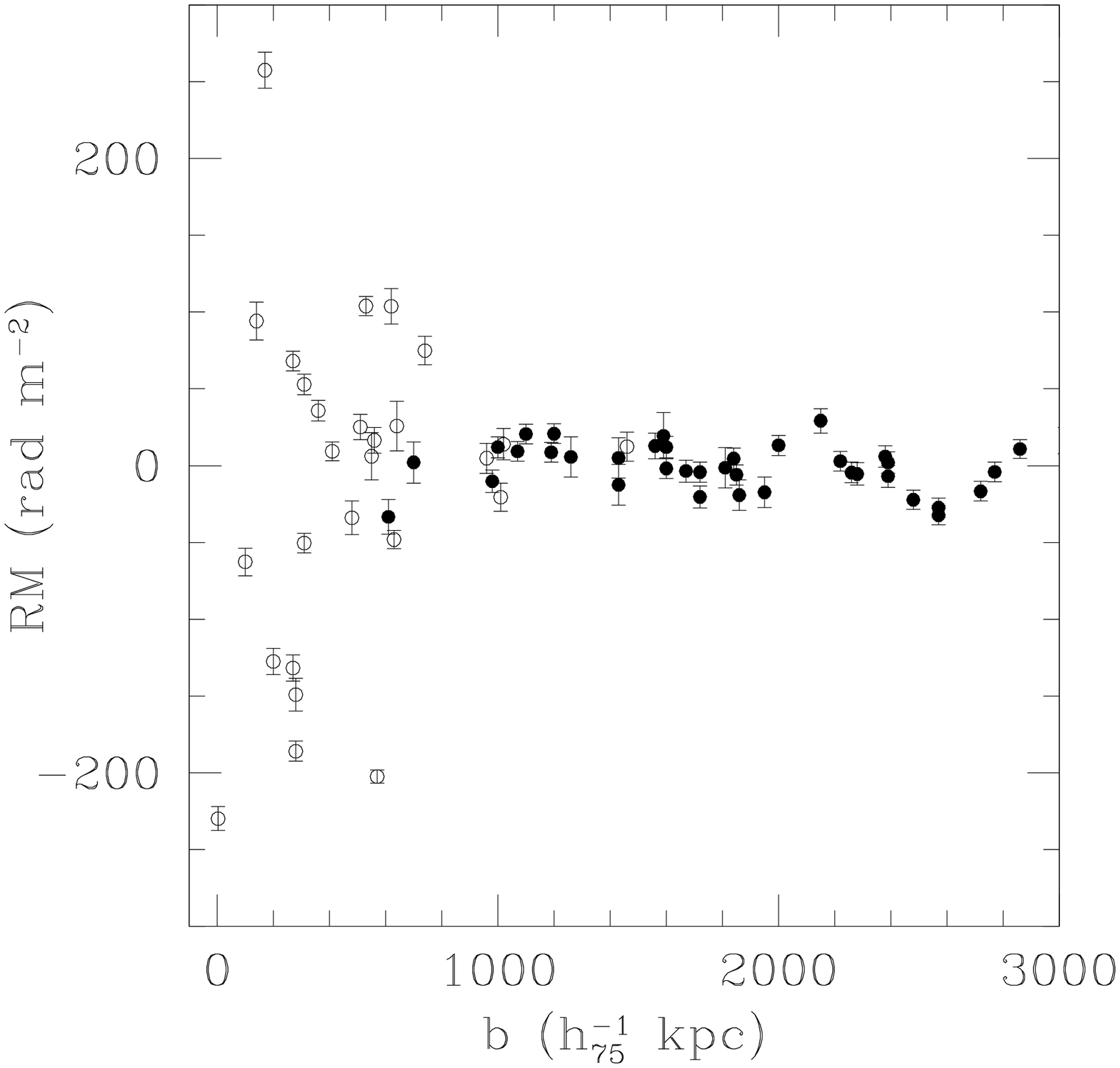}

\figcaption[figure1.ps]{Galaxy-corrected rotation measure plotted as a
function of source impact parameter in kiloparsecs for the sample of
16 Abell clusters. The open points represent the {\it cluster} sources
viewed through the thermal cluster gas while the closed points are the
{\it control} sources at impact parameters beyond the cluster
gas. Note the clear increase in the width of the RM distribution
toward smaller impact parameter.\label{RRM}}


\begin{thebibliography}{}
\bibitem[B\"ohringer (1995)]{hans95} B\"ohringer, H.\ 1995, \revma, 8,
295 
\bibitem[Bowyer, Bergh\"ofer, \& Korpela (1999)]{bowy1} Bowyer, S., Bergh\"ofer, T.W. \& Korpela, E.\ 1999, in {\it Diffuse Thermal and Relativistic Plasma in Galaxy Clusters}, eds. H.\ B\"ohringer, L.\ Feretti, \& P.\ Schuecker, MPE Report 271, 201
\bibitem[Colgate \& Li (1999)]{colgate99} Colgate, S.A. \&
Li, H.\ 1999, in {\it Highly Energetic Physical Processes, and Mechanisms
for Emision from Astrophysical Plasmas}, eds. P.C.H. Martens \&
S. Tsuruta. ASP Conf. Series 334, 255
\bibitem[NVSS, Condon et al.\ (1998)]{nvss} 
Condon, J.J., Cotton, W.D., Greisen, E.W., Yin, Q.F., Perley, R.A., 
Taylor, G.B., \& Broderick, J.J.\ 1998, \aj, 115, 1693
\bibitem[Dogiel (1999)]{dogiel1} Dogiel, V.A. \ 1999, in {\it Diffuse Thermal and Relativistic Plasma in Galaxy Clusters}, eds. H.\ B\"ohringer, L.\ Feretti, \& P.\ Schuecker, MPE Report 271, 259
\bibitem[XBACs, Ebeling et al.\ (1996)]{xbacs} Ebeling, H., Voges, W.,
Bohringer, H., Edge, A.C., Huchra, J.P., \& Briel, U.G.\ 1996, \mnras,
281, 799
\bibitem[En\ss lin, Lieu, \& Biermann (1999)]{ensslin1} En\ss lin, T.A., Lieu, R. \& Biermann, P.L. \ 1999, \aap, 344, 409
\bibitem[Feretti et al.\ (1995)]{feretti95} Feretti, L., Dallacasa,
D., Giovannini, G., \& Tagliani, A.\ 1995, \aap, 302, 680
\bibitem[Goldshmidt \& Rephaeli (1993)]{gr93} Goldshmidt, O.\ \&
Rephaeli, Y.\ 1993, \apj, 411, 518
\bibitem[Hennessy, Owen, \& Eilek (1989)]{hoe89} Hennessy, G.S., Owen,
F.N., \& Eilek, J.A.\ 1989, \apj, 347, 144
\bibitem[Kim et al.\ (1990)]{kk90}
Kim, K.-T., Kronberg, P.P., Dewdney, P.D., \& Landecker, T.L.\ 1990, \apj, 355, 29
\bibitem[Kim et al.\ (1991)]{kim91}
Kim, K.-T., Tribble, P.C., \& Kronberg, P.P.\ 1991, \apj, 379, 80
\bibitem[Lawler \& Dennison(1982)]{ld82} Lawler, J.M.\ \& Dennison,
B.\ 1982, \apj, 252, 81
\bibitem[Rephaeli, Ulmer, \& Gruber(1994)]{rug94} Rephaeli, Y., Ulmer,
M., \& Gruber, D.\ 1994, \apj, 429, 554
\bibitem[Sarazin(1986)]{sarazin86} Sarazin, C.L.\ 1986, \revm, 58, 1
\bibitem[Sarazin \& Kempner (2000)]{sarazin00} Sarazin, C.L., \& Kempner, J.C.\ 2000, \apj, 533, 73
\bibitem[Taylor, Allen, \& Fabian (1999)]{tab99} Taylor, G.B., Allen,
S.W., \& Fabian, A.C.\ 1999, in {\it Diffuse Thermal and Relativistic
Plasma in Galaxy Clusters}, eds. H.\ B\"ohringer, L.\ Feretti, \& P.\
Schuecker, MPE Report 271, 77
\bibitem[V\"olk \& Atoyan(2000)]{voelk00} V\"olk, H.J. \& Atoyan, A.M.\ 2000, \apj, 541, 88
\bibitem[EXSAS, Zimmermann et al.\ (1994)]{exsas} Zimmermann, H.U., Becker,
W., Belloni, T., D\"obereiner, S., Izzo, C., Kahabka, P., \&
Schwentker, O.\ 1994, EXSAS User's Guide, MPE Report 244
\end{thebibliography}
\end{document}